\documentclass[letterpaper, 10 pt, journal, twoside]{template}
\IEEEoverridecommandlockouts                              

\usepackage{graphics} 
\usepackage{epsfig} 
\usepackage{mathptmx} 
\usepackage{times} 
\usepackage{amsmath} 
\usepackage{amssymb}  
\usepackage{color}
\usepackage{booktabs} 
\usepackage{multirow} 
\usepackage{tabularx} 

\usepackage{subfig}   
\usepackage{float}    
\usepackage{enumitem}

\usepackage{makecell}
\usepackage{pifont}
\usepackage{orcidlink}  

\makeatletter
\let\NAT@parse\undefined
\makeatother

\usepackage{booktabs} %
\usepackage{verbatim} %
\usepackage{graphicx} 
\usepackage[numbers,sort, compress]{natbib}
\usepackage{flushend}

\usepackage{hyperref} 
\hypersetup{
	colorlinks=true, 
	linkcolor=blue,  
	citecolor=blue,  
	urlcolor=blue    
}

\usepackage{array}
\newcolumntype{Y}{>{\raggedright\arraybackslash}X}
\newcolumntype{Z}{>{\centering\arraybackslash}X}
\newcolumntype{W}{>{\raggedleft\arraybackslash}X}
\renewcommand{\arraystretch}{1.3} 

\usepackage{framed}   
\definecolor{shadecolor}{rgb}{0.92,0.95,0.95}
\renewcommand{\thetable}{\arabic{table}}

\definecolor{mygrey}{gray}{0.965}
\definecolor{mycyan}{rgb}{0.92,0.95,0.95}

\usepackage{fancyhdr}
\title{Practical Process Capability Indices Workflows} 

\author{
	Fei Jiang\thanks{Independent researchers, Seattle, WA, USA.
		Corresponding author: Fei Jiang, email: jiangfeicq@gmail.com.},
	Lei Yang\footnotemark[1]
}

\begin{document}
\maketitle
\thispagestyle{fancy}
\pagestyle{plain} 
\begin{abstract}
This paper presents a comprehensive review of univariate process capability indices (PCIs), which are critical metrics for assessing how effectively a manufacturing process satisfies customer specifications based on a single quality characteristic. The primary objective of this review is to develop practical procedural workflows for conducting process capability analysis under various preconditions, including those less frequently addressed scenarios in existing literature. Key analytical components, such as outlier detection, normality test, and best distribution fitting, are integrated into the proposed framework to ensure accurate and robust capability assessments. By systematically evaluating a range of methodologies, this study offers guidance for researchers and practitioners in selecting the most appropriate PCIs for specific process conditions. Ultimately, the work aims to simplify the complexity of PCI analysis while enhancing its precision and utility in quality control and process improvement efforts.
\end{abstract}

\section{INTRODUCTION}
\IEEEPARstart{T}{he} process capability is a measurable property of an in-control process to the specification, expressed as a process capability indices (PCIs) \cite{spiring1995process, pearn1992distributional, kotz2002process, anis2008basic}, which are widely used to validate how well a process can produce output within specified limits in the manufacturing industry, offering a quantitative evaluation of a process's ability to produce items that meet the predetermined quality standards. Various PCIs have been introduced to assess both the potential and performance of a process. Notably, widely utilized indices such as $C_p$, $C_{pk}$ $P_p$ and $P_{pk}$ are frequently employed for this purpose \cite{palmer1999review, chen2001process, boyles1991taguchi, wooluru2015views}. $C_p$ and $C_{pk}$ are indices that assess short-term process capability when the process is stable, while $P_p$ and $P_{pk}$ provide insight into long-term performance of the process over time, including all sources of variation. $C_p$, also known as the process capability index, measures the potential capability of a process assuming it is centered within the specification limits \cite{juran1979quality, herman1989capability}, accounting for process centering and evaluates the capability by considering the worst-case scenario relative to the specification limits \cite{taguchi1985quality, kane1986process}. $P_p$, or process performance index, is similar to $C_p$ but is calculated using overall (long-term) standard deviation ($\sigma_{overall}$) to replace $C_p$'s short-term within standard deviaiton ($\sigma_{within}$), making it a more comprehensive measure of actual process performance. $C_{pk}$ extends this concept further by incorporating process centering, offering a more accurate view of actual performance \cite{herman1989capability}. It is calculated by dividing the minimum distance from the mean to the nearest spec limit by three times the standard deviation. Higher $C_{pk}$ values (e.g., $\geq 1.33$) indicate a more capable and reliable process, helping reduce defects and improve quality. $P_{pk}$ is analogous to $C_{pk}$ but using the overall standard deviation instead. Together, these indices offer valuable insights into both immediate and sustained process capability, helping organizations monitor, control, and improve manufacturing quality.

$C_{pm}$ \cite{hsiang1985tutorial, chan1988optimal} and $C_{pmk}$ \cite{pearn1992distributional, sadeghpour2014estimation} are process capability indices that account for deviations from the target value ($T$), making them more sensitive to process centering than $C_p$ and $C_{pk}$. The $C_{pm}$ and $C_{pmk}$ indices are much less known and used less frequently. $C_{pm}$, also referred to as the Taguchi capability index \cite{boyles1991taguchi}, differs from traditional capability indices in that it explicitly incorporates the deviation from the target value. This deviation, or bias, is defined as the difference between the process mean and the target value. Unlike conventional indices that focus solely on conformance to specification limits, $C_{pm}$ introduces a penalty for variation away from the target, thereby encouraging processes that are not only within specifications but also centered around the desired target value \cite{chan1988new, abbasi2016class, pearn2002estimating, sy1995asymptotic}. $C_{pmk}$ extends this concept by also considering the worst-case performance relative to the specification limits, offering a more comprehensive measure of process capability. Similarly, $P_{pm}$ and $P_{pmk}$ are long-term counterparts of $C_{pm}$ and $C_{pmk}$, incorporating overall process performance over extended periods. These indices are particularly useful in industries where meeting target specifications is as critical as maintaining consistency within specification limits. By integrating target-centered capability measures, $C_{pm}$, $C_{pmk}$, $P_{pm}$, and $P_{pmk}$ provide a more holistic assessment of manufacturing quality and process stability \cite{palmer1999review, kotz2002process}.

To further refine process capability assessments, extended indices such as $C_{p}^\ast$, $C_{pk}^\ast$, $C_{pm}^\ast$, $C_{pmk}^\ast$ have been introduced \cite{pearn1992distributional, pearn2000estimating, pearn2002estimating, chang2008assessing, anis2008basic, chen1998incapability, chen2001capability}, incorporating additional corrections for bilateral asymmetric tolerances. $C_{p}^\ast$ adjusts $C_p$ to account target value, ensuring a more accurate measure of potential capability. $C_{pk}^\ast$ extends $C_{pk}$ by integrating skewness corrections, making it more reliable for asymmetric processes. $C_{pm}^\ast$ refines $C_{pm}$ by incorporating advanced loss functions to better quantify deviations from target values, while $C_{pmk}^\ast$ combines the benefits of $C_{pk}^\ast$ and $C_{pm}^\ast$, offering a comprehensive measure that considers both target deviations and worst-case performance. These indices are particularly useful in high-precision industries where traditional PCIs may not fully capture process complexities, enabling a more accurate and robust capability assessment.

New process capability indices such as $C_{Np}$, $C_{Npk}$, $C_{Npm}$ and $C_{Npmk}$ have been developed to address situations where the underlying process distribution is unknown or does not conform to common parametric normal assumptions \cite{chen1997application, chen2001new, anis2008basica, kovarik2014process}. When the distribution is unknown, nonparametric percentiles can be used to calculate PCIs. Alternatively, if a normality test fails, parametric percentiles from the best-fitted distribution can be employed. The generalized non-normal PCIs extend traditional indices to better assess processes with non-normal distributions. $C_{Np}$ serves as a non-normal counterpart to $C_p$, measuring process potential by comparing the specification range to the range between the 0.135th and 99.865th percentiles, replacing the $\pm 3\sigma$ process spread. $C_{Npk}$ accounts for process centering by comparing the distance from the process median to each specification limit, relative to the percentile spread. $C_{Npm}$  adds sensitivity to how close the process median is to the target value by including the squared deviation from the target in the denominator. $C_{Npmk}$ combines the concepts of $C_{Npk}$ and $C_{Npm}$ to evaluate both off-center and off-target performance, offering the most comprehensive view of process capability in non-normal conditions. These indices are particularly valuable for industries dealing with complex or highly variable processes where traditional parametric indices may fail to provide an accurate assessment of process capability.

Additionally, other creative methodologies have been developed to proceed the capability analysis. Vannman introduces a unified class of process capability indices, $C_p(u, v)$ \cite{vannman1995unified}, which generalizes existing indices like $C_p$, $C_{pk}$, $C_{pm}$, and $C_{pmk}$ by varying two parameters. It derives the statistical properties of two estimators under normality and a centered target, showing that while higher sensitivity (larger $u$ and $v$) improves detection of mean shifts, it can lead to poor estimator performance. The study concludes that both the index and its estimator's behavior, especially bias and mean square error, must be considered when choosing a capability measure. This generalized indices is also used in in non-normal distributions, denoted as $C_{Np}(u,v)$ \cite{chen1997application}. Greenwich \textit{et. al} introduced $C_{pp}$, which is equal to the square of the reciprocal of $C_{pm}$ \cite{greenwich1995process, chen1998incapability}. $C_{pp}$ transforms the traditional $C_{pm}$ index to separately measure process inaccuracy and imprecision, offering clearer insights into the causes of process incapability. $C_{pp}$ equals zero for a perfect process and increases with greater deviation from the target or higher variability.

Despite existing research on the calculation of process capability indices, two critical issues remain underexplored, hindering the attainment of more straightforward and accurate results. The first challenge lies in the vast number of scenarios, making it extremely difficult to select the most appropriate method, especially since some scenarios have yet to be explored. Addressing this issue would enhance convenience for engineers in their daily research and work. The second challenge is simplifying the overall process by eliminating non-critical factors or disregarding rare situations, thereby significantly improving work efficiency. This paper aims to tackle these issues by introducing practical workflows that offer clear guidelines for conducting univariate process capability analysis with greater precision and effectiveness. 

The subsequent sections of this paper are organized as follows. Section \ref{section-wf} presents the PCI workflow across different scenarios and introduces methods for outlier detection and normality testing prior to PCI calculation. Section \ref{section-cpk-normal} discusses the various PCIs applicable to normally distributed data, while Section \ref{section-cpk-nonnormal} focuses on PCIs for non-normally distributed data. Section \ref{section-sd} explores how to calculate both within and overall standard deviations under different scenarios using a practical workflow. Finally, section \ref{section-case} validates the proposed workflows through a case study using real engineering data.

\section{Process Capability Indices Workflow}
\label{section-wf}
\begin{figure}[htbp]
	\centerline{\includegraphics[width=1.00\linewidth]{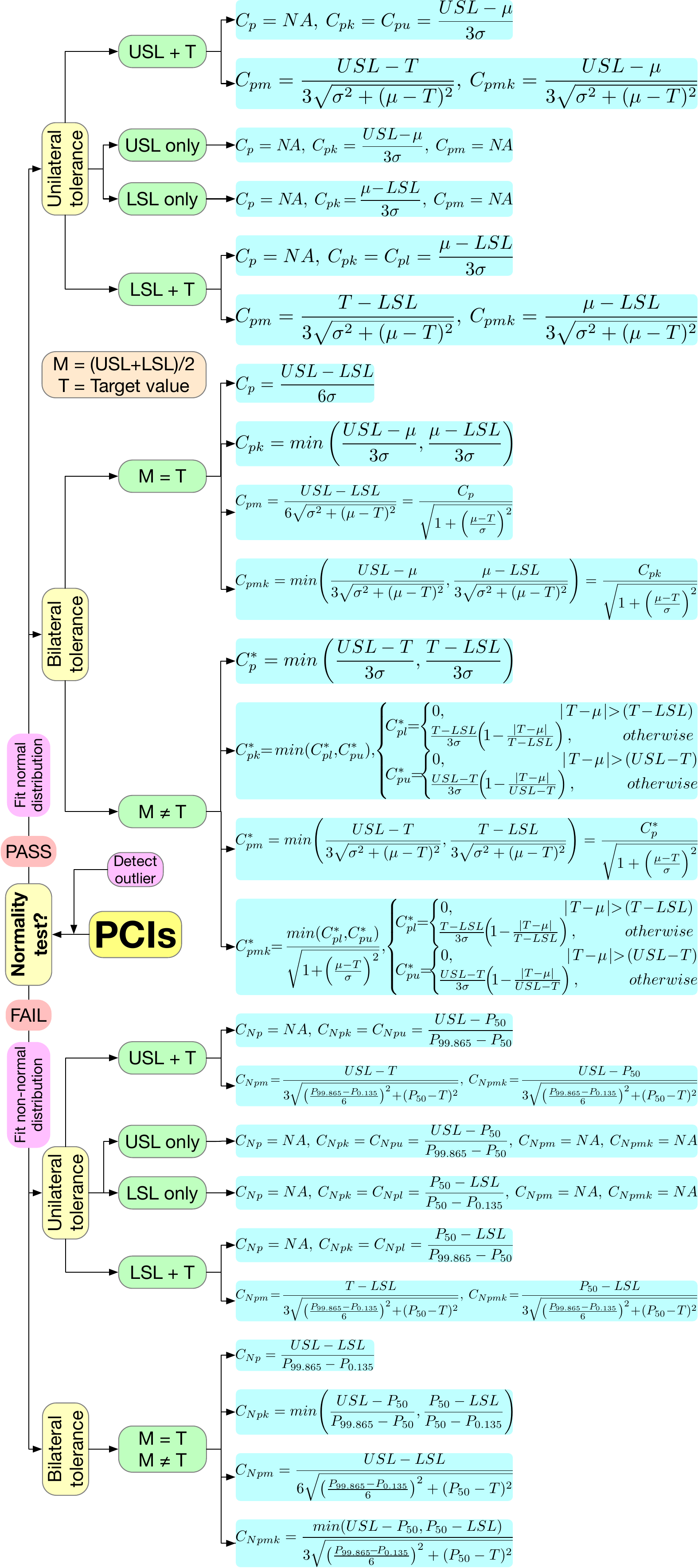}}
	\caption{Practical workflow for calculating various PCIs across different scenarios} 
	\label{cpkwf}
\end{figure}

Figure \ref{cpkwf} illustrates a structured workflow designed to guide the selection of the most appropriate PCIs based on specific predefined conditions. The decision-making process considers several critical factors, such as the type of tolerance (unilateral or bilateral), the presence of outliers, the results of normality testing, best-fitted distribution, and whether a specific target value is defined and aligned with the midpoint of the tolerance range. The workflow systematically addresses these elements, and offers a data-driven approach to selecting the most appropriate PCI, as explored in Sections \ref{section-cpk-normal} and \ref{section-cpk-nonnormal}, thereby enabling manufacturers and quality control professionals to make informed decisions that optimize process performance and product quality.

The workflow begins with outlier detection to identify and address anomalies that could distort analysis results Additionally, the workflow integrates a normality test to ensure that the data distribution aligns with the statistical assumptions underlying PCI calculations, thereby enhancing the accuracy and reliability of process capability evaluations. Outlier detection and the normality test are discussed in the following two subsections. 

\subsection{Normality test}
A normality test evaluates whether a dataset follows a normal distribution, a common assumption in many statistical analyses, by comparing the observed data to the expected normal distribution and producing a p-value that indicates the likelihood of the data being normally distributed \cite{yazici2007comparison, yap2011comparisons}. If the p-value exceeds the chosen significance level $\alpha$ (e.g., $\alpha = 0.05$), the null hypothesis of normality is not rejected, suggesting no significant deviation from normality; if the p-value is less than or equal to $\alpha$, the null hypothesis is rejected, indicating the data is not normally distributed. Common normality test methods include the Shapiro-Wilk (S-W) test \cite{dudley2023shapiro}, which is highly sensitive and ideal for small to moderate sample sizes; the Kolmogorov-Smirnov (K-S) test \cite{berger2014kolmogorov}, which is less powerful but compares the entire distribution; and the Anderson-Darling (A-D) test \cite{nelson1998anderson}, which improves on K-S by giving more weight to the tails. Additionally, the Chi-square normality test compares observed and expected frequencies in binned data to assess fit to a normal distribution, best suited for large samples but sensitive to binning choices. However, as for small sample sizes, the S-K and A-D tests are most appropriate, with S-W generally performing better for normality assessment. However, the A-D test is preferred for process capability analysis and distribution fitting because it emphasizes tail deviations and supports a broader range of distributions, unlike S-W, which is limited to normal and exponential distributions.

If the given dataset passes the normality test, normal distribution-based methodologies are used to calculate PCIs, as discussed in Section \ref{section-cpk-normal}. Otherwise, non-normal distribution-based methodologies are applied, as shown in Section \ref{section-cpk-nonnormal}.

\subsection{Best distribution fitting}
\label{section-fitting}

When a normality test fails, indicating that the data do not follow a normal distribution, it becomes essential to explore alternative probability distributions that better represent the underlying data. Best distribution fitting involves systematically comparing a range of candidate distributions, such as weibull, gamma, log-normal, exponential, or others, to identify the one that most closely aligns with the data. This process typically includes visual assessments (e.g., P-P plots, Q-Q plots, histograms with fitted density curves), statistical goodness-of-fit tests (like the S-W, K-S, A-D and Chi-square tests), Akaike/Bayesian information criteria (AIC/BIC) to quantify how well each distribution fits \cite{chakrabarti2011aic}. By identifying the most suitable distribution, we can improve the accuracy of subsequent statistical analyses, modeling, and decision-making processes that rely on understanding the data’s true probabilistic nature.

Among all the methods, AIC and BIC are often preferred because they are easy to interpret and provide a quantitative way to compare and select distributions. The equations for AIC and BIC are given as follows:
\begin{equation}
	\begin{split}
	AIC = - 2ln(L) + 2k \\
	BIC = - 2ln(L) + k \times log(n) \\
	AICc = - 2ln(L) + 2k + \frac{2k(k+1)}{n-k-1}
	\end{split}
\end{equation}
where $L$ represents the maximum value of the likelihood function for the model, k denotes the number of parameters in the selected distribution (e.g., $k=1$ for exponential, $k=2$ for normal, and $k=3$ for a 3-parameter weibull), and $n$ represents the sample size.

AIC and BIC are commonly used to evaluate model fit, with smaller values indicating better models, though both criteria increase with the number of parameters, potentially penalizing more complex models. Since information criteria are not hypothesis tests, the distribution with the minimum AIC may still be a very poor fit. For smaller sample sizes, AICc is preferred over AIC due to its correction for finite samples. However, there is no universal consensus on which criterion is best, and opinions vary. Additionally, many practitioners attempt to identify the best-fitting distribution by selecting the one with the highest p-value from tests like S-W and A-D. However, this practice is inappropriate. It is important to understand that while a small p-value may suggest rejecting a distribution, a large p-value does not confirm that the data follows it.

\section{PCIs for Normal Distributed Data}
\label{section-cpk-normal}
The methodologies for calculating PCIs vary depending on whether the data follows a normal or non-normal distribution, and selecting an inappropriate PCI can lead to inaccurate conclusions and poor decision-making. This section focuses on the correct application of PCIs for normally distributed data, providing a detailed explanation of key indices such as $C_p$, $C_{pk}$, $P_p$, $P_{pk}$, $C_{pm}$ and $C_{pmk}$. These indices will be discussed under different preconditions, including unilateral tolerances both with and without a specified target value, as well as bilateral tolerances with either symmetric or asymmetric limits. Understanding the appropriate use of each index in relation to the distribution and tolerance conditions is essential to accurately evaluating a process's capability and ensuring valid data-based decisions.

\subsection{The $C_p$ index}
\label{section-cp}
The $C_p$ index is a statistical measure that quantifies a process’s ability to produce output within specified limits, assuming the process is centered within those limits. It is calculated as the ratio of the tolerance range ($T_r = USL - LSL$) to six times the process standard deviation ($6\sigma$), representing the natural spread of the process. A higher $C_p$ value indicates a more capable process. When $C_p  \in (-\infty, 1,00)$ , it typically means that a process is not capable of consistently producing output within the specified limits; $C_p  \in [1,00, 1,33)$ suggests a marginal capability while $C_p  \in [1.33, +\infty)$ indicates the process is capable, with low variation and a strong ability to consistently produce within specification limits. However, the $C_p$ index does not account for how well the process is centered. To provide a more complete evaluation, indices like $C_{pk}$ are used, as they consider potential sample mean shifts in the process.

The $C_p$ index varies for different scenarios:
\subsubsection{Bilateral symmetric tolerance ($M=T$)}
\begin{equation}
	C_p = \frac{USL-LSL}{6\sigma}
	\label{cp}
\end{equation}
where $USL$ represents the upper specification limit, and $LSL$ represents the lower specification limit. The midpoint of the specification range, denoted as $M$, is calculated as $M=(USL+LSL)/2$. $T$ denotes the target value, also known as the nominal value. The within standard deviation is represented by $\sigma$, denoted specifically as $\sigma_{within}$, which is used in the calculations of short-term PCIs, including $C_p$, $C_{pl}$, $C_{pu}$, $C_{pk}$, $C_{pm}$ and $C_{pmk}$ (see Section \ref{section-cp} to \ref{section-cpmk}). The overall standard deviaion, denoted as $\sigma_{overall}$, is used for long-term PCIs like $ P_p$ and $P_{pk}$ (see Section \ref{section-ppk}). The detailed methodologies of standard deviation are discussed in Section \ref{section-sd}.

Bilateral tolerance defines the permissible variation in a dimension both above and below its nominal value, allowing deviations in both directions. This means that neither the upper nor the lower tolerance is zero. It can be symmetrical, where the tolerance is equal in both directions ($M=T$) or asymmetrical ($M \neq T$), with different upper and lower limits. This type of tolerance is used in engineering design and verificaiton to ensure flexibility in manufacturing processes while maintaining the functional accuracy of the part.

\subsubsection{Bilateral asymmetric tolerance ($M \neq T$)}
\begin{equation}
	C_p^\ast = min \left( \frac{USL - T}{3\sigma}, \frac{T - LSL}{3\sigma} \right)
	\label{cpstar}
\end{equation}
where $C_p^\ast $ denotes the $C_p $ index under the assumption that $M \neq T$. The same applies to  $C_{pk}^\ast$, $C_{pm}^\ast$ and $C_{pmk}^\ast$ as shown in Equations (\ref{cpkast}), (\ref{cpmast}) and (\ref{cpmkast}), respectively,.

\subsubsection{Unilateral tolerance}
The $C_p$ index measures process capability based on bilateral tolerances, where both upper and lower tolerances are non-zero. However, in cases of unilateral tolerances, where only an $USL$ or $LSL$ is defined, or when either upper or lower tolerance equals zero, the $C_p$ index loses its meaning or becomes undefined. This limitation is particularly relevant in geometric dimensioning and tolerancing (GD\&T) applications, where many features, such as position, flatness, and surface profile, are controlled using unilateral tolerance zones \cite{liu2013process}. As a result, alternative capability indices or evaluation methods are often needed for these types of features.

\subsection{The $C_{pk}$ index}
\label{section-cpk}
The $C_{pk}$ assess how well a process produces outputs within specified limits while considering both process variation and centering. It quantifies the ability of a process to meet specifications by comparing the spread of data to the allowable tolerance range, with higher $C_{pk}$ values indicating a more capable and consistent process. A $C_{pk}$ greater than 1.33 is generally considered acceptable, while lower values suggest potential issues with variation or misalignment with target specifications. Widely used in industries such as electronics, automotive, aerospace and advanced manufacturing, $C_{pk}$ helps organizations optimize processes, reduce defects, and improve product quality. 

The $C_{pk}$ index also varies for different scenarios:

\subsubsection{Bilateral symmetric tolerance ($M=T$)}
\begin{equation}
	C_{pk} = min\left(\frac{USL - \mu}{3\sigma},\frac{\mu - LSL}{3\sigma}\right)
	\label{cpk}
\end{equation}
$C_{pk}$ represents the minimum distance from the process mean to either the $USL$ or $LSL$, scaled by three times the standard deviation ($3\sigma$). Unlike $C_p$, which assumes the process is perfectly centered, $C_{pk}$ accounts for any shift in the mean. As a result, $C_{pk}$ is typically less than $C_{p}$ unless the process is well-centered.

\subsubsection{Bilateral  asymmetric tolerance ($M \neq T$)}
When the process mean's target value $T$ differs from the midpoint $M$ of specification limits, corresponding indices are defined as follows \cite{anis2008basic}:
\begin{equation}
	\label{cpkast}
	\footnotesize
	C_{pk}^\ast \!\! = \! min(C_{pl}^\ast, \! C_{pu}^\ast), \hspace{-0.2em}
	\left\{ \hspace{-0.75em}
	\begin{array}{lr}
		C_{pl}^\ast \!\! = \!\!
		\left\{
		\begin{array}{lr}
			\hspace{-0.5em} 0, &  \hspace{-1.75em} \mid \! T \! - \! \mu \! \mid > \! (T \! - \! LSL)\\
			\hspace{-0.5em} \! \frac{T-LSL}{3\sigma} \!\! \left( \! 1 \! - \! \frac{\mid T-\mu \mid}{T-LSL} \right), & otherwise
		\end{array}
		\right.
		\\
		C_{pu}^\ast \!\! = \!\! \hspace{-0.2em}
		\left\{
		\begin{array}{lr}
			\hspace{-0.5em} 0, & \hspace{-1.75em} \mid \! T \! - \! \mu \! \mid > \! (USL \! - \!T)\\
			\hspace{-0.5em} \! \frac{USL-T}{3\sigma} \!\! \left( \! 1 \! - \! \frac{\mid T-\mu \mid}{USL-T} \right), & otherwise
		\end{array}
		\right.
	\end{array}
	\right.\\
\end{equation}

However, in most cases, the manufacturing process is designed to center the dimension within the middle of the specified tolerance range, regardless of whether the tolerance is symmetric or asymmetric. As a result, the $C_{pk}$ is not affected by the target value. For simplicity, Equation (\ref{cpk}) can be used in this context instead.

\subsubsection{Unilateral tolerance with $USL$ only}
\begin{equation}
	C_{pk} =C_{pu} = \frac{USL  -  \mu}{3\sigma}
\end{equation}

\subsubsection{Unilateral tolerance with $LSL$ only}
\begin{equation}
	C_{pk} = C_{pl} = \frac{\mu - LSL}{3\sigma}
\end{equation}

\subsection{The $C_{pm}$ index}
\label{section-cpm}
The $C_{pm}$ index is an advanced statistical measure used in quality control to evaluate how well a process meets target specifications while considering both process variation and deviation from the target value. Unlike $C_{pk}$, which only accounts for process spread and centering within specification limits, $C_{pm}$ incorporates the target value, making it a more accurate representation of process performance when aiming for a specific optimal value. Thus, a target value must be specified to calculate the $C_{pm}$. A higher $C_{pm}$ indicates a process that is not only capable of staying within limits but also consistently aligned with the desired target, reducing variation and improving overall product quality. This index is particularly valuable in industries where achieving precise target values is crucial, such as semiconductor manufacturing, pharmaceuticals, and precision engineering.

\subsubsection{Bilateral symmetric tolerance ($M=T$)}
\begin{equation}
	{C}_{{pm}}={\frac  {USL-LSL}{6 \sqrt{{\sigma }^2 + (\mu-T)^2}}}={\frac  {{C}_{p}}{{\sqrt  {1+\left({\frac  {{\mu }-T}{\sigma}}\right)^{2}}}}}
\end{equation}

\subsubsection{Bilateral symmetric tolerance ($M \neq T$)}
\begin{equation}
	\label{cpmast}
	{C}_{pm}^\ast= \frac{min(USL - T, T - LSL)}{3\sqrt{{\sigma }^2 + (\mu-T)^2}} = {\frac  {{C}_{p}^\ast}{{\sqrt  {1+\left({\frac  {{\mu }-T}{\sigma}}\right)^{2}}}}}
\end{equation}
where $C_p^\ast$ is calcuated in Equation (\ref{cpstar}).

\subsubsection{Unilateral tolerance with $USL$ and target specified}
\begin{equation}
	C_{pm} = \frac{USL-T}{3\sqrt{{\sigma }^2 + (\mu-T)^2}}
\end{equation}

\subsubsection{Unilateral tolerance with $LSL$ and target specified}
\begin{equation}
	C_{pm} = \frac{T-LSL}{3\sqrt{{\sigma }^2 + (\mu-T)^2}}
\end{equation}

\subsubsection{Unilateral tolerance with $USL$ or $USL$ only}
The $C_{pm}$ index assumes the presence of a target value, along with either or both upper and lower specification limits. Therefore, it is not well-defined when applied to unilateral tolerances that include only a $USL$ or $LSL$ without a specified target value.

\subsection{The $C_{pmk}$ index}
\label{section-cpmk}
The $C_{pmk}$ index \cite{sy1995asymptotic} is usually used in quality control to assess a process's ability to meet specification limits while considering both process variation and its alignment with the target value. It is an extension of $C_{pm}$, incorporating adjustments for process centering relative to specification limits, making it a more comprehensive indicator of process performance. Unlike $C_{pk}$, which only considers the nearest specification limit, $C_{pmk}$ accounts for deviations from the optimal target, providing a more accurate representation of true process capability. A higher $C_{pmk}$ value indicates a well-controlled process with minimal variation and closer adherence to target specifications, making it particularly valuable in high-precision industries like electronics, pharmaceuticals, and aerospace manufacturing.

\subsubsection{Bilateral symmetric tolerance ($M=T$)}
\begin{equation}
	{C}_{{pmk}}= \frac{min(USL-\mu, \mu-T)}{3\sqrt{\sigma^2 + (\mu-T)^2}} ={\frac  {{C}_{{pk}}}{{\sqrt  {1+\left({\frac  {{\mu }-T}{\sigma}}\right)^{2}}}}}
\end{equation}

\subsubsection{Bilateral asymmetric tolerance ($M \neq T$)}
\
\begin{equation}
	\footnotesize
	\label{cpmkast}
	\begin{split}
		C_{pmk}^\ast \!\! = \! \frac{min(C_{pl}^\ast, \! C_{pu}^\ast)}{\sqrt{1 \! + \! \left( \frac{\mu - T}{\sigma} \right)^2 }}, 
		\left\{ \hspace{-0.75em}
		\begin{array}{lr}
			C_{pl}^\ast \!\! = \!\!
			\left\{
			\begin{array}{lr}
				\hspace{-0.5em} 0, &  \hspace{-1.75em} \mid \! T \! - \! \mu \! \mid > \! (T \! - \! LSL)\\
				\hspace{-0.5em} \! \frac{T-LSL}{3\sigma} \!\! \left( \! 1 \! - \! \frac{\mid T-\mu \mid}{T-LSL} \right), & otherwise
			\end{array}
			\right.
			\\
			C_{pu}^\ast \!\! = \!\! \hspace{-0.2em}
			\left\{
			\begin{array}{lr}
				\hspace{-0.5em} 0, & \hspace{-1.75em} \mid \! T \! - \! \mu \! \mid > \! (USL \! - \!T)\\
				\hspace{-0.5em} \! \frac{USL-T}{3\sigma} \!\! \left( \! 1 \! - \! \frac{\mid T-\mu \mid}{USL-T} \right), & otherwise
			\end{array}
			\right.
		\end{array}
		\right.
	\end{split}
\end{equation}

\subsubsection{Unilateral tolerance with $USL$ and $T$ specified}
\begin{equation}
	C_{pmk} = \frac{USL-\mu}{3\sqrt{{\sigma }^2 + (\mu-T)^2}}
\end{equation}

\subsubsection{Unilateral tolerance with $LSL$ and $T$ specified}
\begin{equation}
	C_{pmk} = \frac{\mu - LSL}{3\sqrt{{\sigma }^2 + (\mu-T)^2}}
\end{equation}

\subsubsection{Unilateral tolerance without $T$ specified}
The $C_{pmk}$ index is not well-defined when applied to unilateral tolerances, where only a $USL$ or $LSL$ is provided without a specified target value.

\subsection{The $P_p$ and $P_{pk}$ indices}
\label{section-ppk}
Process performance indices ($P_p$ and $P_{pk}$) and process capability indices ($C_p$ and $C_{pk}$) both assess how well a process meets specification limits, but they differ in their purpose and how they calculate standard deviation. $C_p$ and $C_{pk}$ measure a process’s potential capability when it is in statistical control, using the within (short-term) standard deviation, which isolates natural process variation. $P_p$ and $P_{pk}$, however, evaluate the actual performance of the process over time, using the overall (long-term) standard deviation, which includes both natural variation and any shifts or trends in the process. This distinction means that $C_p$ and $C_{pk}$ are more useful for predicting consistent long-term performance when a process is stable, while $P_p$ and $P_{pk}$ use overall standard deviation, providing a snapshot of current process performance, even if it is unstable, such as the metrology data from prototyping or engineering verification builds. Since long-term variation is typically larger than short-term variation, $P_p$ and $P_{pk}$ values are often lower than $C_p$ and $C_{pk}$, highlighting the impact of process shifts and inconsistencies. The equations of $P_p$ and $P_{pk}$ are shown below.
\begin{equation}
	P_p = \frac{USL - LSL}{6\sigma_{overall}}
	\label{pp}
\end{equation}

\begin{equation}
	\begin{split}
		{P}_{{pu}}= \frac{USL - \mu}{3\sigma_{overall}}, \quad {P}_{{pl}}= \frac{\mu - LSL}{3\sigma_{overall}} \\
		{P}_{{pk}}= min\left(P_{pl}, P_{pu} \right) = min\left(\frac{USL-\mu}{3\sigma_{overall}}, \frac{\mu-LSL}{3\sigma_{overall}} \right)
		\label{ppk}
	\end{split}
\end{equation}
Equations (\ref{pp}) and (\ref{ppk}) assume bilateral symmetric tolerances. For cases involving bilateral asymmetric or unilateral tolerances, a similar approach applies by replacing the within standard deviation with the overall standard deviation in Section \ref{section-cp} and \ref{section-cpk}.

Likewise, the long-term $C_{pm}$ and $C_{pmk}$ indices can also be computed by substituting $\sigma_{within}$ with $\sigma_{overall}$ in the equations presented in Sections \ref{section-cpm} and \ref{section-cpmk}.

\section{PCIs for Non-normal Data}
\label{section-cpk-nonnormal}

Calculating $C_{pk}$ using methods based on the assumption of normality when dealing with non-normal data is fundamentally incorrect, yet it’s surprisingly common. It’s essential to recognize that not all data follow a normal distribution. Many natural processes, like waiting times or time-to-failure, often follow distributions such as the exponential, which are skewed and not symmetric. When faced with non-normal data, you have two options: one is to transform the data using methods like Box-Cox or Johnson transformations to approximate normality \cite{atkinson2021box}, allowing PCIs to be calculated using the normal based methodologies, though this alters the original data and can cause confusion in interpreting results. The better option is to identify and use a distribution that accurately represents your data both theoretically and practically, preserving the data’s context and improving process understanding. This approach avoids calculating PCIs altogether since it relies on assumptions that no longer hold. The critical first step is determining the correct distribution (see Section \ref{section-fitting}), which can be done using software tools, but the chosen distribution must also make logical sense for the process being studied; for example, while a gamma distribution might provide a better statistical fit for certain data, an exponential distribution may be more appropriate based on the actual nature of the process. 

Many industrial processes do not follow a normal distribution, such as GD\&T dimensions with unilateral tolerances \cite{liu2013process}, rendering conventional PCIs inadequate for accurately assessing process capability. To overcome this limitation, researchers have developed generalized PCIs that incorporate percentile-based estimations and alternative statistical methods suited for non-normal data \cite{chen1997application}. Modified indices like $C_{Npk}$ and its variations use robust measures such as medians and percentiles instead of the traditional mean and standard deviation \cite{chen2001new}. These advancements allow for more accurate capability assessments across diverse manufacturing environments, supporting better decision-making in quality improvement and process optimization.

\begin{figure}[htbp]
	\centerline{\includegraphics[width=1.00\linewidth]{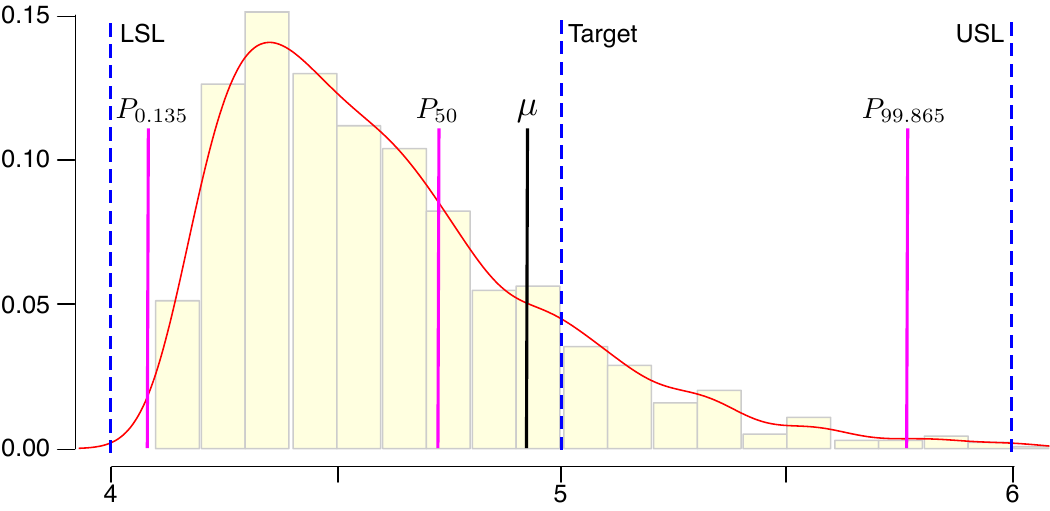}}
	\caption{PCIs-related parameters for a non-normally distributed data, with histogram and density plot integrated} 
	\label{non-normal}
\end{figure}

Figure \ref{non-normal} presents a schematic of non-normally distributed data, combining a histogram and a density plot. Key parameters are displayed in the plot, including the $USL$, $LSL$, target value, sample mean, and three selected percentiles ($P_{0.135}$, $P_{50}$ and $P_{99.865}$), which will be used in Equations (\ref{cnp1})-(\ref{cnpmk1}).

\subsection{Bilaterial Symmetric Tolerance}
The non-normal PCIs are estimated as \cite{kovarik2014process}:
\begin{equation}
	{C}_{Np}=\frac{USL-LSL}{P_{99.865} - P_{0.135}}
	\label{cnp1}
\end{equation}

\begin{equation}
	{C}_{{Npk}}=min {\Bigg (}{USL-{P_{50} } \over {P_{99.865} - P_{50}}},{ {P_{50} }-LSL \over {P_{50} - P_{0.135}}}{\Bigg )}
\end{equation}

\begin{equation}
	{C}_{Npm}=\frac{USL-LSL}{6\sqrt{ \left( \frac{P_{99.865} - P_{0.135}}{6} \right)^2 + (P_{50} - T)^2 }} 
\end{equation}

\begin{equation}
	{C}_{{Npmk}} = \frac{min (USL - P_{50}, P_{50} - LSL)}{3\sqrt{ \left( \frac{P_{99.865} \! - \!  P_{0.135}}{6} \right)^2 + (P_{50} - T)^2}}
\end{equation}
where letter ``N" represents ``non-normal" distribution, while $P_{99.863}$ and $P_{0.135}$ are the 99.863 and 0.135 percentiles of the selected distribution. Similarly, $P_{50}$ is the 50 percentile, also know as ``median" value ($\widetilde{M}$). The use of these percentiles is justified to mimic the normal based PCIs to cover $\pm 3\sigma$ ($99.73\%$) of the data. In situations where the data follows a normal distribution, $P_{50}$ can be replaced with the mean $\mu$, and range ($P_{99.865} - P_{0.135}$) can be replaced with $6\sigma$.

\subsection{Unilateral tolerance}
The $C_p$ index is not well-defined for all unilateral tolerances, whereas the $C_{pm}$ index is undefined for unilateral tolerances with only a $USL$ or $LSL$ without a specified target value.
\subsubsection{Unilateral tolerance, with $USL$ only}
\begin{equation}
	C_{Npk} = C_{Npu} = \frac{USL - P_{50}}{P_{99.865} - P_{50}}
	\label{nnormcpk1}
\end{equation}

\subsubsection{Unilateral tolerance, with $LSL$ only}
\begin{equation}
	C_{Npk} = C_{Npl} = \frac{P_{50} - LSL}{P_{50} - P_{0.135}}
	\label{nnormcpk2}
\end{equation}

\subsubsection{Unilateral tolerance, $USL$ with target specified}
\begin{equation}
	C_{Npm} \! = \! \frac{USL - T}{3\sqrt{ \left( \frac{P_{99.865} - P_{0.135}}{6} \right)^2 \! + \! (P_{50} \! - \! T)^2 }}
\end{equation}
\begin{equation}
	{C}_{{Npmk}} = \frac{USL - P_{50}}{3\sqrt{ \left( \frac{P_{99.865} \! - \!  P_{0.135}}{6} \right)^2 + (P_{50} - T)^2}}
\end{equation}

$C_{Npk}$ is equal to Equation (\ref{nnormcpk1}).

\subsubsection{Unilateral tolerance, $USL$ with target specified}
\begin{equation}
	C_{Npm} \! = \! \frac{T - LSL}{3\sqrt{ \left( \frac{P_{99.865} - P_{0.135}}{6} \right)^2 \! + \! (P_{50} \! - \! T)^2 }}
\end{equation}
\begin{equation}
	C_{Npmk} \! = \! \frac{P_{50} - LSL}{3\sqrt{ \left( \frac{P_{99.865} - P_{0.135}}{6} \right)^2 \! + \! (P_{50} \! - \! T)^2 }}
	\label{cnpmk1}
\end{equation}
$C_{Npk}$ is equal to Equation (\ref{nnormcpk2}).

Traditional normal PCIs can produce misleading results when applied to non-normal data, as they often underestimate or overestimate true capability due to skewness or heavy tails in the distribution. Those non-normal capability methods are more accurately reflect actual process performance by incorporating the true distribution shape, particularly in the tails, making them essential for reliable quality assessment when the assumption of normality is violated.

\section{Standard Deviation Workflow}
\label{section-sd}
Those PCIs are heavily impacted by the standard deviation. The primary distinction between short-term and long-term PCIs lies in how the standard deviation is estimated \cite{mahmoud2010estimating, alvarez2015methodological}. In quality control applications, different estimation methods are used to determine within and overall standard deviation based on sample characteristics and available data. 

The within standard deviation, commonly applied in process control and capability analysis, is typically estimated using within subgroup variation, adjusted by a control chart constant as a correction factor. These estimators are particularly useful when multiple independent small-sized samples are collected over time.

The overall standard deviation captures both within and between variations, providing a comprehensive measure of variability in a process over time. It accounts for all sources of variation, including short-term fluctuations, long-term drifts, and shifts in the process mean. In contrast, within standard deviation is derived from smaller subgroups and focuses on short-term process stability. Because overall standard deviation considers the entire dataset, it reflects both inherent process variation and external influences. This broader perspective makes it crucial for assessing a process's true long-term capability and is essential for performance indices such as $P_p$ and $P_{pk}$, which evaluate overall process stability and quality consistency. 

\begin{figure}[htbp]
	\centerline{\includegraphics[width=1.00\linewidth]{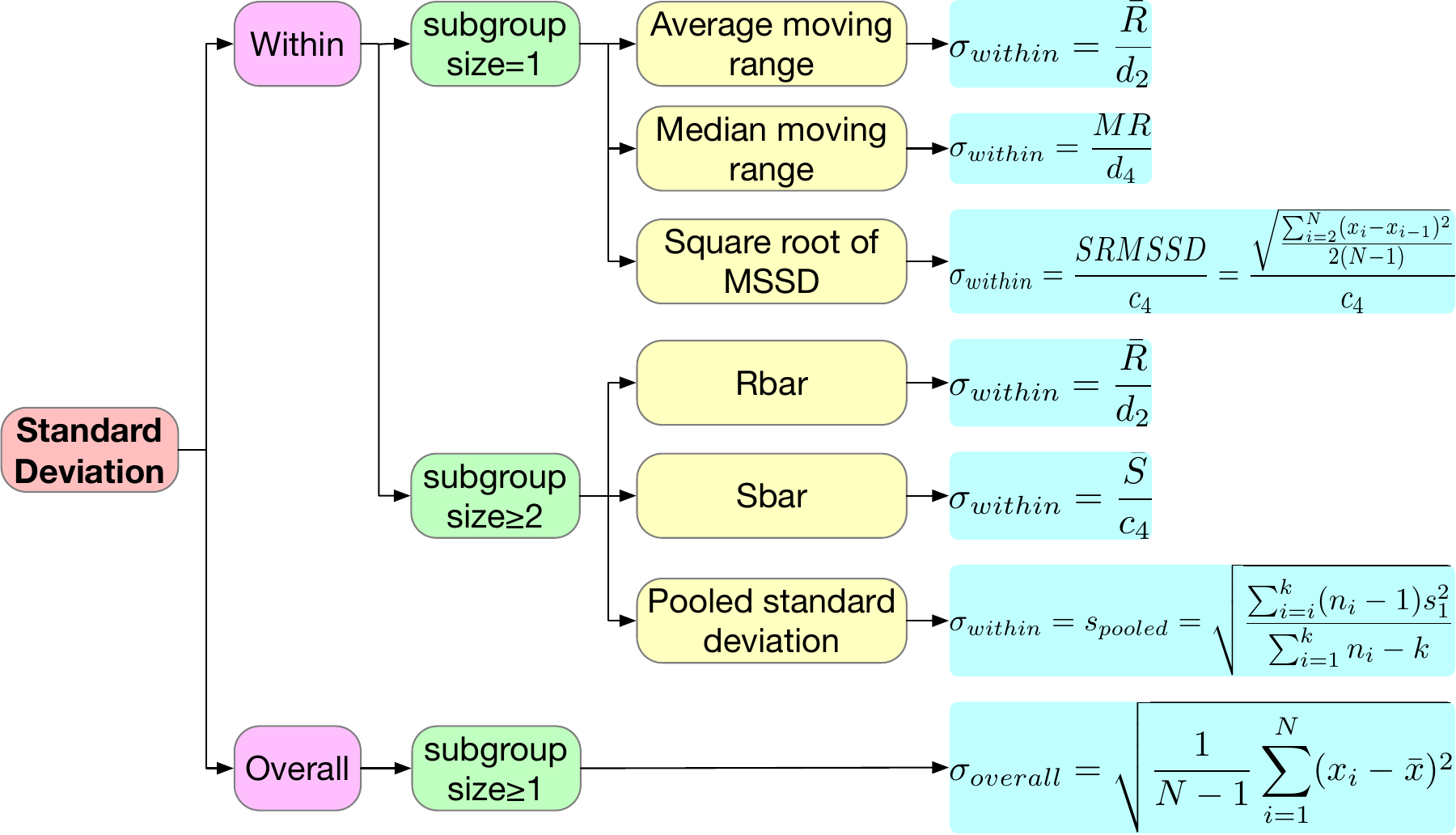}}
	\caption{Standard deviation estimation workflow} 
	\label{sd1}
\end{figure}

Figure \ref{sd1} presents a practical workflow outlining the step-by-step process for calculating both within standard deviaiton and overall standard deviation using different methodologies, which will be illustrate in the following sections.

\subsection{Within Standard Deviation ($\sigma_{within}$)}
Within standard deviation is a key measure of short-term process variability, and its calculation depends on subgroup size and data characteristics. To ensure accuracy, the data must be arranged in temporal order.

For subgroup size = 1, three common methods are used: (1) Average Moving Range (AMR), which estimates variability based on the average of consecutive sample differences; (2) Median Moving Range (MMR), which uses the median of these differences for a more robust estimation; and (3) Square Root of the Mean Squared Successive Differences (SQRT-MSSD), which calculates variability from squared differences between consecutive observations. 

For subgroup size $\geq$ 2, three methods are applied: (1) Rbar ($\overline{R}$), which estimates within standard deviation using the average range of subgroups adjusted by a control chart constant; (2) Sbar ($\overline{S}$), which relies on the average standard deviation of subgroups; and (3) Pooled Standard Deviation, which combines standard deviations from multiple subgroups for a more stable estimate. 

The choice of method depends on the available data and process monitoring needs to ensure an accurate assessment of short-term variation \cite{mahmoud2010estimating}. 
\subsubsection{Average Moving Range}
\index{Average Moving Range}
\index{$N_{subgroup}$}
\index{$\sigma_{xbar}$}
\index{$\overline{R}$}
\index{$Rbar$}
\index{$d_2(w)$}
\index{$w$}

AMR can estimate standard deviation when data consists of individual observations with subgroup size of one, time-ordered observations from a stable process, by analyzing variability between consecutive data points. It calculates the average of these moving ranges and divides it by a constant (see Table \ref{tab:constnt1}) to obtain the approximate within standard deviation, which is used to calculate $C_p$, $C_{pk}$, $C_{pu}$, $C_{pk}$, $C_{pm}$ and $C_{pmk}$. Unlike the overall standard deviation, change the order of the data can significantly affect the $\sigma_{within}$. AMR is commonly used in statistical process control and ideal for monitoring stability and detecting trends in sequential data with limited samples.

\begin{equation}
	\begin{split}
		\sigma_{within} = \frac{\overline{MR}}{d_2(w)}, \\
		\overline{MR} = \frac{MR_w + ... MR_i ... + MR_n}{n-w+1} = \frac{\sum_{i=w}^{n}MR_i}{n-w+1}, \\
		MR_i = max \left( x_i,...,x_{i-w+1} \right) - min\left( x_i,...,x_{i-w+1} \right), \\
		i=w,...,n.
	\end{split}
\end{equation}

where $w$ is represents the number of observations used in the moving range (the default value is $w=2$); $n$ is the total number of samples; $(n-w+1)$  denotes the number of subgroups, which is equal to the number of moving ranges; $MR_i$ is the $i^{th}$ moving range, where each subgroup contains w samples; $\overline{MR}$ is the mean of $\sum_{i=w}^{n}MR_i$; $d_2(w)$ is an unbiased constant for control chart that depends on the number of sampes  in each subgroup, obtained from table \ref{tab:constnt1} \cite{oakland2007statistical}.

\begin{table}[htbp]
	\centering
	\footnotesize
	\renewcommand{\arraystretch}{1.0} 
	\setlength{\tabcolsep}{12pt} 
	\begin{tabular}{ccccc}
		\toprule
		\makecell{Sample \\ size ($w$)} & \textbf{$d_2$} & \textbf{$c_4$} & \textbf{$d_3$} & \textbf{$d_4$} \\
		\midrule
		2     & 1.1284 & 0.7979 & 0.8525 & 0.9539 \\
		3     & 1.6926 & 0.8862 & 0.8884 & 1.5878 \\
		4     & 2.0588 & 0.9213 & 0.8798 & 1.9783 \\
		5     & 2.3259 & 0.9400 & 0.8641 & 2.2569 \\
		6     & 2.5344 & 0.9515 & 0.8480 & 2.4717 \\
		7     & 2.7044 & 0.9594 & 0.8332 & 2.6455 \\
		8     & 2.8472 & 0.9650 & 0.8198 & 2.7908 \\
		9     & 2.9700 & 0.9693 & 0.8078 & 2.9154 \\
		10    & 3.0775 & 0.9727 & 0.7971 & 3.0242 \\
		\bottomrule
	\end{tabular}
	\caption{Constants used in control charts for sample sizes from 2 to 10}
	\label{tab:constnt1}
\end{table}

The most commonly used moving range size is 2, where the range is calculated between two consecutive observations. This is standard practice in individuals ($X$) control charts, as it provides a simple yet effective measure of short-term variability. Larger moving range sizes can be used but may reduce sensitivity to small shifts in the process.

\index{Control Chart Constants}

\subsubsection{Median of  Moving Range}
\index{Median of  Moving Range}
\index{$\sigma_{xbar} $}
\index{$\overline{MR}$}
\index{$d_4(w)$}

The MMR method is similar as AMR, except that the average moving range ($\overline{R}$) is replaced with the median moving range ($\widetilde{MR}$).

\begin{equation}
	\begin{split}
		\sigma_{within} = \frac{\widetilde{MR}}{d_4(w)}, \\
		MR_i = max \left( x_i,...,x_{i-w+1} \right) - min\left( x_i,...,x_{i-w+1} \right), \\
		i=w,...,n.
	\end{split}
\end{equation}

where $w$ is represents the number of observations used in the moving range (the a default value is $w=2$); n is the total number of samples; $(n-w+1)$  denotes the number of subgroups, which is equal to the number of moving ranges; $MR_i$ is the $i^{th}$ moving range, where each subgroup contains w samples; $\widetilde{MR}$ is the median of sorted $MR_i$ values; $d_4(w)$ is an unbiased control chart constant read from table \ref{tab:constnt1}.

\index{$MR_i$}
\index{$w$}

\subsubsection{Square root of MSSD (SRMSSD)}
\index{$MSSD$}

MSSD (mean squared successive differences) stands for the mean of squared successive differences. The square root of MSSD is the square root of the mean of the squared differences between consecutive points. Use this method when you cannot reasonably assume that at least 2 consecutive points were collected under similar conditions \cite{w1993estimating}.

SRMSSD has two primary applications: in basic statistics, it is used to test whether a sequence of observations is random by comparing the estimated population variance with the Mean Square Successive Difference (MSSD); in control charts, it serves as an estimator for variance when the subgroup size is 1.

For cases when you can't assume that two successive points form a rational subgroup nor use the moving range methods, the MSSD method provides an alternative. The estimation of within standard deviation is calculated as,

\index{$SRMSSD$}

\begin{equation}
	\sigma_{within} = SRMSSD = \frac{\sqrt{\frac{\sum_{i=2}^{N} (x_i - x_{i-1})^2}{2(N-1)}}}{c_4(w)}
\end{equation}

where $N$ represents the total number of observations;  $w$ is  the number of observations in $i^{th}$ subgroup; $c_4(w)$ is the unbiasing constant from Table \ref{tab:constnt1}. 

\index{$MR_i$}
\index{$c_4(n_i)$}
\index{$c_4'(n_i)$}
\index{$N$}
\index{$n_i$}

\subsubsection{Average of subgroup ranges (Rbar, $\overline{R}$)}

\index{$Rbar$}
\index{$\overline{R}$}
\index{$N_{subgroup}$}
\index{$d_2$}

It's a classical method for estimating standard deviation when the subgroup size is greater than or equal to 2. The average range is calculated as the mean of the subgroup ranges. The within standard deviation is then estimated from the following equation:

\begin{equation}
	\sigma_{within} = \frac{\overline{R}}{d_2} = \frac{\sum R_i / k}{d_2}
\end{equation}

where $R_i$ is the range of the $i^{th}$ subgroup, $k$ is the number of subgroups, and $d_2$ is a constant that depends on subgroup size.

\subsubsection{Average of subgroup standard deviation (Sbar, $\overline{S}$)}

\index{$Sbar$}
\index{$\overline{S}$}
\index{$c_4$}

This method is similar as the average of subgroup ranges, but instead of $\overline{R}$, it utilizes $\overline{s}$ along wiht the corresponding constant $c_4$.  When the subgroup size is constant, the average of the subgroup standard deviations is given by:

\begin{equation}
	\sigma_{within} = \frac{\overline{s}}{c_4} = \frac{\sum s_i / k}{d_2}
\end{equation}
where $s_i$ is the standard deviation of the $i^{th}$ subgroup , $k$ is the number of subgroups, and $c_4$ is a constant that depends on subgroup size.

\subsubsection{Pooled Standard Deviation}

\index{Pooled Standard Deviation}
\index{$S_{pooled}$}
\index{$n_1$}
\index{$n_2$}
\index{$s_1$}
\index{$s_2$}

The pooled standard deviation is a statistical measure that combines the standard deviations of multiple groups to estimate a common variability when assuming equal variances \cite{rudmin2010calculating}. It is particularly useful in hypothesis testing, such as t-tests, ANOVAs,  effect size calculations, where comparing group means requires a stable estimate of dispersion. They are also used in lab-based sciences like biology and chemistry, where they can be an indication for repeatability of an experiment. By weighting each sample’s variance based on its size, the pooled standard deviation provides a more reliable measure than a simple average of individual standard deviations. This approach ensures a fair comparison across groups, making it a fundamental tool in data analysis where homogeneity of variance is assumed.

The within standard deviaiton equals to pooled standard deviation ($s_{pooled}$), which is calculated using the following formula:

\begin{equation}
	\sigma_{within} = \sqrt{\frac{\sum_{i=i}^{k}(n_i-1)s_i^2}{\sum_{i=1}^{k}n_i - k}} = \sqrt{\frac{s_1^2 + s_2^2 + ... + s_k^2}{k}}
\end{equation}
where the standard deviation of each group are represented as $s_1$, $s_2$, ..., $s_k$, while the sample sizes of each group are denoted as $n_1$, $n_2$, ..., $n_k$. The total number of groups is given by k.

For two groups, the formula simplifies to:
\begin{equation}
	S_{pooled} = \sqrt{\frac{(n_1 - 1)s_1^2 + (n_2-1)s_2^2}{n_1 + n_2 - 2}}
\end{equation}
where $n_1$ and $n_2$ are sample size for group 1 and group 2, while $s_1$ and $s_2$ are standard deviation for group 1 and group 2, respectively.

\subsection{Overall Standard Deviation ($\sigma_{overall}$)}

\index{$\sigma_{overall}$}

Overall standard deviation is an estimate of overall standard deviation. It can be used to calculate overall (long-term) process capability indicies such as $Pp$, $Ppu$, $Ppl$ and $Ppk$.
\begin{equation}
	\sigma_{overall}=\sqrt{{\frac{1}{N-1}}\sum_{i=1}^N(x_i-\overline{x})^2}, \:\:\: where \:\:\: \mu=\frac{1}{N}\sum_{i=1}^Nx_i
\end{equation}
where $N$ is the total number of data points.

According to the equation, it adds up the squares of those deviations, dividing  by either $N-1$ (for an unbiased estimate from a sample) or $N$ (for a biased estimate when treating the data as the entire population), and then take the square root.  The calculated standard deviation as the average distance each individual data point is from the overall average.  Note that you use all the data in the calculation.  This is why this standard deviation is sometimes called the overall variation.  It accounts for all the variation in the data.

\section{PCIs Workflow Case Study}
\label{section-case}

\subsection{Standard deviation comparison}
This section will utilize normally distributed sample datasets with symmetric tolerance (see Table \ref{tab:rawdata}), to illustrate the importance of selecting the appropriate PCIs during data review and analysis. The dataset has a subgroup size of one, encompassing nine dimensions ($101 \sim 109$) and 32 samples for each, with a specified target value ($T$), upper tolerance ($Tol+$), and lower tolerance ($Tol-$). The analysis will include a comparison between overall standard deviation and within standard deviation. Additionally, it will examine the differences between key capability indices, specifically $C_p$ versus $P_p$ and $C_{pk}$ versus $P_{pk}$, to highlight their implications in process evaluation. To enhance clarity, visual representations such as plots will be incorporated to illustrate these comparisons and their impact on data interpretation.

\begin{table}[htbp]
	\centering
	\footnotesize
	\renewcommand{\arraystretch}{1.0}
	\setlength{\tabcolsep}{1.65pt}
	\begin{tabular}{cccccccccc}
		\toprule
		NO. & 101 & 102 & 103 & 104 & 105 & 106 & 107 & 108 & 109 \\
		\midrule
		$T$ & 4.620 & 9.060 & 10.780 & 17.000 & 23.580 & 28.530 & 34.750 & 42.600 & 50.780 \\
		$Tol+$ & 0.100 & 0.100 & 0.100 & 0.100 & 0.100 & 0.100 & 0.100 & 0.100 & 0.100 \\
		$Tol-$ & 0.100 & 0.100 & 0.100 & 0.100 & 0.100 & 0.100 & 0.100 & 0.100 & 0.100 \\
		\midrule
		1  & 4.650 & 8.979 & 10.711 & 16.963 & 23.632 & 28.473 & 34.754 & 42.619 & 50.777 \\
		2  & 4.636 & 8.958 & 10.735 & 17.008 & 23.601 & 28.490 & 34.736 & 42.643 & 50.753 \\
		3  & 4.658 & 9.002 & 10.761 & 17.030 & 23.605 & 28.572 & 34.871 & 42.645 & 50.774 \\
		4  & 4.662 & 8.938 & 10.770 & 16.998 & 23.629 & 28.507 & 34.667 & 42.601 & 50.764 \\
		5  & 4.662 & 9.025 & 10.759 & 16.997 & 23.614 & 28.489 & 34.756 & 42.648 & 50.762 \\
		6  & 4.662 & 8.972 & 10.738 & 16.976 & 23.611 & 28.519 & 34.683 & 42.621 & 50.751 \\
		7  & 4.646 & 9.002 & 10.753 & 17.003 & 23.641 & 28.529 & 34.783 & 42.640 & 50.777 \\
		8  & 4.669 & 9.006 & 10.761 & 16.996 & 23.610 & 28.459 & 34.734 & 42.648 & 50.758 \\
		9  & 4.641 & 9.014 & 10.754 & 16.972 & 23.625 & 28.508 & 34.751 & 42.661 & 50.795 \\
		10 & 4.644 & 8.964 & 10.776 & 17.005 & 23.619 & 28.518 & 34.782 & 42.640 & 50.813 \\
		11 & 4.623 & 8.924 & 10.769 & 17.056 & 23.599 & 28.589 & 34.794 & 42.611 & 50.761 \\
		12 & 4.632 & 8.954 & 10.765 & 16.968 & 23.613 & 28.516 & 34.704 & 42.637 & 50.757 \\
		13 & 4.654 & 9.034 & 10.733 & 16.984 & 23.629 & 28.553 & 34.808 & 42.610 & 50.804 \\
		14 & 4.661 & 9.002 & 10.768 & 16.991 & 23.625 & 28.499 & 34.766 & 42.642 & 50.773 \\
		15 & 4.638 & 9.030 & 10.753 & 16.992 & 23.623 & 28.520 & 34.814 & 42.662 & 50.804 \\
		16 & 4.629 & 8.957 & 10.744 & 17.001 & 23.620 & 28.494 & 34.769 & 42.630 & 50.751 \\
		17 & 4.605 & 9.052 & 10.736 & 16.981 & 23.604 & 28.501 & 34.778 & 42.623 & 50.782 \\
		18 & 4.626 & 9.016 & 10.760 & 17.002 & 23.608 & 28.507 & 34.885 & 42.615 & 50.806 \\
		19 & 4.610 & 8.917 & 10.779 & 17.012 & 23.651 & 28.542 & 34.697 & 42.616 & 50.747 \\
		20 & 4.623 & 9.023 & 10.755 & 16.990 & 23.617 & 28.532 & 34.804 & 42.614 & 50.827 \\
		21 & 4.623 & 8.941 & 10.776 & 17.020 & 23.622 & 28.536 & 34.803 & 42.605 & 50.760 \\
		22 & 4.629 & 9.028 & 10.760 & 16.995 & 23.640 & 28.598 & 34.900 & 42.622 & 50.788 \\
		23 & 4.675 & 9.045 & 10.729 & 16.985 & 23.641 & 28.546 & 34.868 & 42.652 & 50.801 \\
		24 & 4.640 & 8.978 & 10.750 & 16.943 & 23.604 & 28.460 & 34.642 & 42.646 & 50.727 \\
		25 & 4.610 & 8.912 & 10.808 & 17.012 & 23.591 & 28.555 & 34.734 & 42.571 & 50.763 \\
		26 & 4.651 & 8.974 & 10.726 & 16.951 & 23.594 & 28.462 & 34.604 & 42.637 & 50.716 \\
		27 & 4.632 & 8.911 & 10.802 & 17.024 & 23.616 & 28.485 & 34.748 & 42.642 & 50.783 \\
		28 & 4.629 & 8.937 & 10.766 & 16.990 & 23.616 & 28.543 & 34.716 & 42.609 & 50.815 \\
		29 & 4.622 & 9.005 & 10.754 & 16.982 & 23.606 & 28.527 & 34.781 & 42.628 & 50.779 \\
		30 & 4.689 & 8.904 & 10.682 & 16.931 & 23.635 & 28.557 & 34.842 & 42.688 & 50.668 \\
		31 & 4.644 & 9.016 & 10.749 & 16.956 & 23.588 & 28.506 & 34.779 & 42.625 & 50.734 \\
		32 & 4.646 & 8.943 & 10.764 & 17.026 & 23.615 & 28.592 & 34.823 & 42.625 & 50.768 \\
		\bottomrule
	\end{tabular}
	\caption{Normally distributed dataset with bilateral symmetric tolerances}
	\label{tab:rawdata}
\end{table}

\begin{table*}[h]
	\centering
	\footnotesize
	\renewcommand{\arraystretch}{1.0} 
	\setlength{\tabcolsep}{1.25pt} 
	\begin{tabular}{cccccccccccccccccccc}
		\toprule
		NO. & $\sigma_{overall}$ & $\sigma_{w.A2}$ & $\sigma_{w.A3}$ & $\sigma_{w.A4}$ & $\sigma_{w.A5}$ & $\sigma_{w.A6}$ & $\sigma_{w.A7}$ & $\sigma_{w.A8}$ & $\sigma_{w.A9}$ & $\sigma_{w.A10}$ & $\sigma_{w.M2}$ & $\sigma_{w.M3}$ & $\sigma_{w.M4}$ & $\sigma_{w.M5}$ & $\sigma_{w.M6}$ & $\sigma_{w.M7}$ & $\sigma_{w.M8}$ & $\sigma_{w.M9}$ & $\sigma_{w.M10}$ \\
		\midrule
		101 & 0.0197 & 0.0165 & 0.0172 & 0.0178 & 0.0181 & 0.0186 & 0.0191 & 0.0193 & 0.0194 & 0.0193 & 0.0168 & 0.0157 & 0.0157 & 0.0175 & 0.0186 & 0.0212 & 0.0201 & 0.0192 & 0.0185 \\
		102 & 0.0434 & 0.0516 & 0.0472 & 0.0457 & 0.0440 & 0.0434 & 0.0427 & 0.0422 & 0.0414 & 0.0410 & 0.0660 & 0.0548 & 0.0480 & 0.0454 & 0.0445 & 0.0423 & 0.0459 & 0.0458 & 0.0443 \\
		103 & 0.0241 & 0.0236 & 0.0237 & 0.0233 & 0.0226 & 0.0225 & 0.0219 & 0.0214 & 0.0210 & 0.0208 & 0.0220 & 0.0211 & 0.0182 & 0.0191 & 0.0174 & 0.0163 & 0.0165 & 0.0158 & 0.0152 \\
		104 & 0.0264 & 0.0273 & 0.0265 & 0.0259 & 0.0258 & 0.0255 & 0.0255 & 0.0257 & 0.0257 & 0.0258 & 0.0262 & 0.0208 & 0.0273 & 0.0301 & 0.0312 & 0.0299 & 0.0290 & 0.0283 & 0.0291 \\
		105 & 0.0153 & 0.0151 & 0.0156 & 0.0152 & 0.0153 & 0.0155 & 0.0156 & 0.0155 & 0.0154 & 0.0154 & 0.0157 & 0.0161 & 0.0152 & 0.0144 & 0.0170 & 0.0172 & 0.0168 & 0.0161 & 0.0155 \\
		106 & 0.0371 & 0.0385 & 0.0369 & 0.0362 & 0.0366 & 0.0378 & 0.0386 & 0.0393 & 0.0395 & 0.0397 & 0.0388 & 0.0403 & 0.0364 & 0.0374 & 0.0384 & 0.0384 & 0.0405 & 0.0446 & 0.0430 \\
		107 & 0.0682 & 0.0702 & 0.0674 & 0.0684 & 0.0696 & 0.0702 & 0.0714 & 0.0702 & 0.0697 & 0.0697 & 0.0681 & 0.0642 & 0.0637 & 0.0711 & 0.0761 & 0.0767 & 0.0727 & 0.0698 & 0.0675 \\
		108 & 0.0219 & 0.0221 & 0.0226 & 0.0223 & 0.0226 & 0.0228 & 0.0230 & 0.0233 & 0.0228 & 0.0224 & 0.0220 & 0.0205 & 0.0238 & 0.0215 & 0.0206 & 0.0197 & 0.0204 & 0.0201 & 0.0198 \\
		109 & 0.0320 & 0.0340 & 0.0333 & 0.0320 & 0.0307 & 0.0295 & 0.0289 & 0.0288 & 0.0287 & 0.0287 & 0.0335 & 0.0334 & 0.0283 & 0.0268 & 0.0251 & 0.0268 & 0.0287 & 0.0274 & 0.0265 \\
		\bottomrule
	\end{tabular}
	\caption{Comparison summary of overall standard deviation and within standard deviation calculated from the dataset in Table \ref{tab:rawdata}. The within standard deviation, is denoted as $\sigma_{w.Ai}$ for AMR method and $\sigma_{w.Mi}$ for MMR method, where i (ranging from 2 to 10) is the number of samples per subgroup.}
	\label{tab:sdvalues}
\end{table*}

\subsubsection{$\sigma_{within}$ versus $\sigma_{overall}$}
Table \ref{tab:sdvalues} presents the calculated values for overall standard deviation and within standard deviation, the latter derived using the AMR and MMR methods. Here, $\sigma_{overall}$ represents the overall standard deviation, while $\sigma_{w.Ai}$  denotes the within standard deviation estimated using the AMR method with a moving range size of i ($i = 2 \sim 10$). Similarly, $\sigma_{w.Mi}$ represents the within standard deviation based on the MMR method with the same moving range size of $i$.

The standard deviation values presented in the Table \ref{tab:sdvalues} reveal that the ratio $\left(\frac{\sigma_{within}}{\sigma_{overall}}\right)$ ranges from 0.631 to 1.521, while $\left(\frac{\sigma_{w.A2}}{\sigma_{overall}}\right)$, where $\sigma_{w.A2}$ represents the commonly used within standard deviation for a subgroup size of one, ranges from 0.838 to 1.189. These wide ranges highlights a substantial discrepancy between the within standard deviation and the overall standard deviation, which impacts the PCIs significantly.

\subsubsection{$\left( \frac{C_p}{P_p} \right)$ and $\left( \frac{C_{pk}}{P_{pk}} \right)$}
\begin{figure}[htbp]
	\centerline{\includegraphics[width=1.00\linewidth]{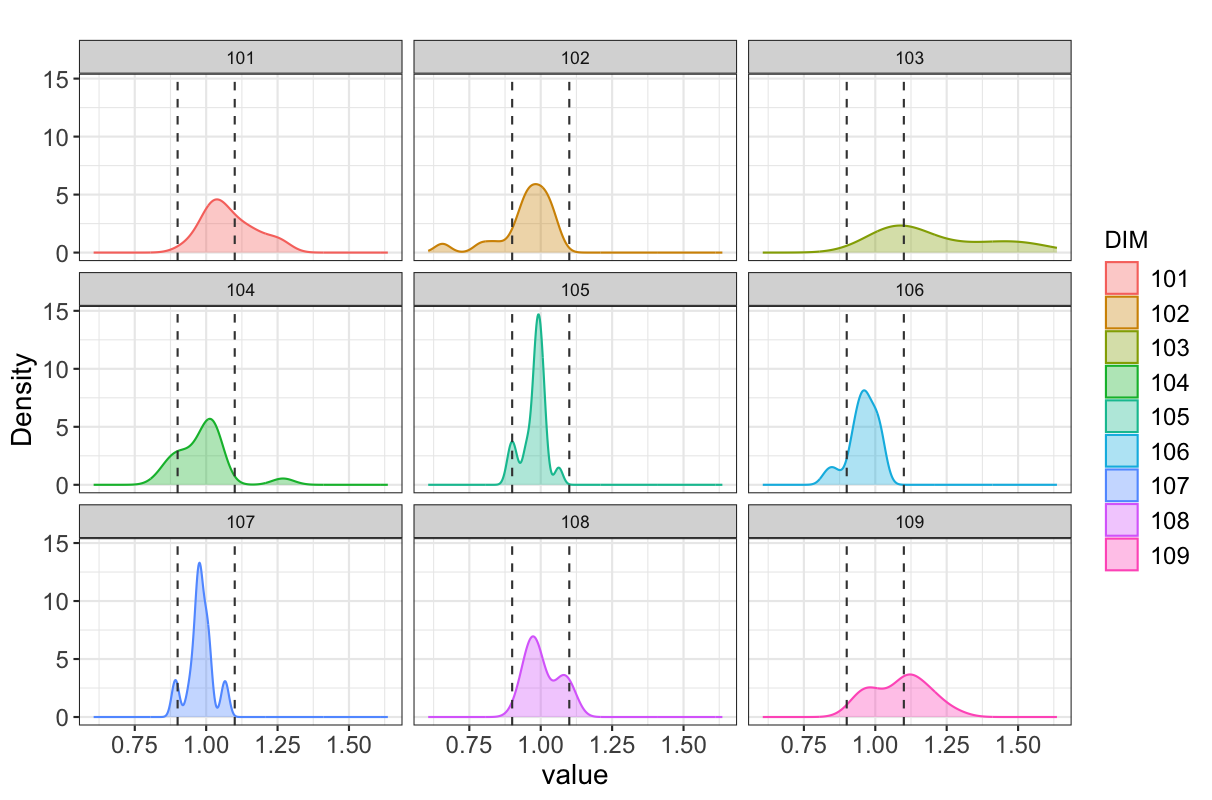}}
	\caption{Histogram with density of $\left(\frac{C_{pk}}{P_{pk}}\right)$ across the 9 dimensions, based on $C_{pk}$ and $P_{pk}$ values in Table \ref{tab:pcis-ppk-vs}, with dashed lines at ratio limits 0.9 and 1.1.}
	\label{non-normal2}
\end{figure}

Based on Equations (\ref{cp}) and (\ref{pp}), the values of $C_p$ and $P_p$ are calculated and presented in Table \ref{tab:pcis-pp-vs}. Similarly, $C_{pk}$ and $P_{pk}$ are derived from Equations (\ref{cpk}) and (\ref{ppk}), with their corresponding values shown in Table \ref{tab:pcis-ppk-vs}. Since $C_p$, $P_p$, $C_{pk}$ and $P_{pk}$ are inversely proportional to the standard deviation, the ratios of $\left( \frac{C_p}{P_p} \right)$ and $\left( \frac{C_{pk}}{P_{pk}} \right)$ equals to $\left( \frac{\sigma_{overall}}{\sigma_{within}} \right)$, falling between 0.657 and 1.585 when all within standard deviations from Table \ref{tab:sdvalues} are considered (see Figure \ref{non-normal2}). If the commonly used $\sigma_{w.A2}$ is used instead, the ratio ranges from 0.841 to 1.193. Regardless of which within-standard deviation is chosen for PCI calculation, the impact on the results is significant. Therefore, careful consideration is essential when selecting the appropriate methodology.
\begin{table*}[h]
	\centering
	\footnotesize
	\renewcommand{\arraystretch}{1.0} 
	\setlength{\tabcolsep}{3.2pt} 
	\begin{tabular}{cccccccccccccccccccc}
		\toprule
		NO. & $P_{p}$ & $C_{p.A2}$ & $C_{p.A3}$ & $C_{p.A4}$ & $C_{p.A5}$ & $C_{p.A6}$ & $C_{p.A7}$ & $C_{p.A8}$ & $C_{p.A9}$ & $C_{p.A10}$ & $C_{p.M2}$ & $C_{p.M3}$ & $C_{p.M4}$ & $C_{p.M5}$ & $C_{p.M6}$ & $C_{p.M7}$ & $C_{p.M8}$ & $C_{p.M9}$ & $C_{p.M10}$ \\
		\midrule
		101 & 1.689 & 2.017 & 1.934 & 1.869 & 1.838 & 1.789 & 1.742 & 1.724 & 1.715 & 1.724 & 1.981 & 2.119 & 2.119 & 1.901 & 1.789 & 1.569 & 1.655 & 1.733 & 1.799 \\
		102 & 0.768 & 0.646 & 0.706 & 0.729 & 0.758 & 0.768 & 0.781 & 0.790 & 0.805 & 0.813 & 0.505 & 0.608 & 0.694 & 0.734 & 0.749 & 0.788 & 0.726 & 0.728 & 0.752 \\
		103 & 1.383 & 1.412 & 1.406 & 1.430 & 1.475 & 1.481 & 1.522 & 1.557 & 1.587 & 1.602 & 1.515 & 1.580 & 1.831 & 1.745 & 1.916 & 2.045 & 2.020 & 2.110 & 2.193 \\
		104 & 1.262 & 1.220 & 1.257 & 1.286 & 1.291 & 1.307 & 1.307 & 1.296 & 1.296 & 1.291 & 1.272 & 1.602 & 1.220 & 1.107 & 1.068 & 1.114 & 1.149 & 1.177 & 1.145 \\
		105 & 2.172 & 2.201 & 2.130 & 2.186 & 2.172 & 2.144 & 2.130 & 2.144 & 2.158 & 2.158 & 2.117 & 2.064 & 2.186 & 2.308 & 1.955 & 1.932 & 1.978 & 2.064 & 2.144 \\
		106 & 0.898 & 0.865 & 0.903 & 0.920 & 0.910 & 0.881 & 0.863 & 0.848 & 0.843 & 0.839 & 0.859 & 0.827 & 0.915 & 0.891 & 0.868 & 0.868 & 0.823 & 0.747 & 0.775 \\
		107 & 0.488 & 0.474 & 0.494 & 0.487 & 0.478 & 0.474 & 0.466 & 0.474 & 0.477 & 0.477 & 0.489 & 0.518 & 0.522 & 0.468 & 0.437 & 0.434 & 0.458 & 0.477 & 0.493 \\
		108 & 1.520 & 1.506 & 1.473 & 1.493 & 1.473 & 1.460 & 1.447 & 1.429 & 1.460 & 1.486 & 1.513 & 1.624 & 1.399 & 1.548 & 1.616 & 1.690 & 1.632 & 1.656 & 1.681 \\
		109 & 1.041 & 0.980 & 1.000 & 1.041 & 1.085 & 1.129 & 1.153 & 1.157 & 1.161 & 1.161 & 0.994 & 0.997 & 1.177 & 1.243 & 1.327 & 1.243 & 1.161 & 1.216 & 1.257 \\
		\bottomrule
	\end{tabular}
	\caption{Comparison of $C_p$ and $P_p$ Values: $C_p$ is based on the within standard deviation, and $P_p$ on the overall standard deviation. Standard deviation values are listed in Table \ref{tab:sdvalues}, with $C_p$ and $P_p$ calculated using Equations (\ref{cp}) and (\ref{pp}), respectively.}
	\label{tab:pcis-pp-vs}
\end{table*}

\begin{table*}[htbp]
	\centering
	\footnotesize
	\renewcommand{\arraystretch}{1.0} 
	\setlength{\tabcolsep}{2.6pt} 
	\begin{tabular}{cccccccccccccccccccc}
		\toprule
		NO. & $P_{pk}$ & $C_{pk.A2}$ & $C_{pk.A3}$ & $C_{pk.A4}$ & $C_{pk.A5}$ & $C_{pk.A6}$ & $C_{pk.A7}$ & $C_{pk.A8}$ & $C_{pk.A9}$ & $C_{pk.A10}$ & $C_{pk.M2}$ & $C_{pk.M3}$ & $C_{pk.M4}$ & $C_{pk.M5}$ & $C_{pk.M6}$ & $C_{pk.M7}$ & $C_{pk.M8}$ & $C_{pk.M9}$ & $C_{pk.M10}$  \\
		\midrule
		101 & 1.329 & 1.587 & 1.522 & 1.471 & 1.446 & 1.408 & 1.371 & 1.357 & 1.350 & 1.357 & 1.558 & 1.668 & 1.668 & 1.496 & 1.408 & 1.235 & 1.303 & 1.364 & 1.415 \\
		102 & 0.154 & 0.130 & 0.142 & 0.146 & 0.152 & 0.154 & 0.157 & 0.158 & 0.161 & 0.163 & 0.101 & 0.122 & 0.139 & 0.147 & 0.150 & 0.158 & 0.146 & 0.146 & 0.151 \\
		103 & 1.031 & 1.053 & 1.048 & 1.066 & 1.099 & 1.104 & 1.135 & 1.161 & 1.183 & 1.195 & 1.129 & 1.178 & 1.365 & 1.301 & 1.428 & 1.524 & 1.506 & 1.573 & 1.635 \\
		104 & 1.159 & 1.121 & 1.155 & 1.181 & 1.186 & 1.200 & 1.200 & 1.191 & 1.191 & 1.186 & 1.168 & 1.471 & 1.121 & 1.017 & 0.981 & 1.023 & 1.055 & 1.081 & 1.051 \\
		105 & 1.368 & 1.386 & 1.342 & 1.377 & 1.368 & 1.350 & 1.342 & 1.350 & 1.359 & 1.359 & 1.333 & 1.300 & 1.377 & 1.454 & 1.231 & 1.217 & 1.246 & 1.300 & 1.350 \\
		106 & 0.820 & 0.790 & 0.824 & 0.840 & 0.831 & 0.805 & 0.788 & 0.774 & 0.770 & 0.766 & 0.784 & 0.755 & 0.836 & 0.813 & 0.792 & 0.792 & 0.751 & 0.682 & 0.707 \\
		107 & 0.401 & 0.390 & 0.406 & 0.400 & 0.393 & 0.390 & 0.383 & 0.390 & 0.392 & 0.392 & 0.402 & 0.426 & 0.429 & 0.385 & 0.359 & 0.357 & 0.376 & 0.392 & 0.405 \\
		108 & 1.057 & 1.047 & 1.024 & 1.038 & 1.024 & 1.015 & 1.006 & 0.993 & 1.015 & 1.033 & 1.052 & 1.129 & 0.973 & 1.077 & 1.124 & 1.175 & 1.135 & 1.152 & 1.169 \\
		109 & 0.936 & 0.881 & 0.899 & 0.936 & 0.976 & 1.015 & 1.036 & 1.040 & 1.044 & 1.044 & 0.894 & 0.897 & 1.058 & 1.118 & 1.193 & 1.118 & 1.044 & 1.093 & 1.130 \\
		\bottomrule
	\end{tabular}
	\caption{Comparison of $C_{pk}$ and $P_{pk}$ Values: $C_{pk}$ is based on the within standard deviation, and $P_{pk}$ on the overall standard deviation. Standard deviation values are listed in Table \ref{tab:sdvalues}, with $C_{pk}$, and $P_{pk}$ calculated using Equations (\ref{cpk}) and (\ref{ppk}), respectively.}
	\label{tab:pcis-ppk-vs}
\end{table*}

\subsection{Standard Deviation Abosolute Relative Error}
Table \ref{tab:ppk_cpk_ratio} presents the calculated within standard deviations, derived using the AMR-based and MMR-based method with moving range sizes ranging from 2 to 10, along with the overall standard deviation for a total of 784 dimensions with 32 samples each. Among these, 489 dimensions passed the normality test. As shown in the left half of the table, the absolute relative difference between the average AMR-based within standard deviaiton ($\sum_{i=2}^{10} \sigma_{w.Ai} / 9$) and overall standard deviations is then computed and expressed in ratio form to quantify their discrepancy, for each category listed in the first column.

\begin{table*}[h]
	\centering
	\footnotesize
	\renewcommand{\arraystretch}{1.0} 
	\setlength{\tabcolsep}{5.0pt} 
	\begin{tabular}{ccccc|ccccc}
		\toprule
		NO. & $\frac{| \frac{\sum_{i=2}^{10} \sigma_{w.Ai} }{9}- \sigma_{overall} |}{\sigma_{overall}} (\%) $ & Ratio & PCT (\%) & PCT.sum (\%) & NO. & $\frac{| \frac{\sum_{i=2}^{10} \sigma_{w.Mi} }{9}- \sigma_{overall} |}{\sigma_{overall}} (\%) $ & Ratio & PCT (\%) & PCT.sum (\%) \\
		\midrule
		1 & [0.0\%, 5.0\%)  & 388$|$784 & 49.49 & 49.49 & 1 & [0.0\%, 5.0\%)      & 260$|$784 & 33.16 & 33.16  \\
		2 & [5.0\%, 7.5\%)  & 131$|$784 & 16.71 & 66.20 & 2 & [5.0\%, 7.5\%)      & 143$|$784 & 18.24 & 51.40 \\
		3 & [7.5\%, 10\%)   &  94$|$784 & 11.99 & 78.19 & 3 & [7.5\%, 10\%)       &  96$|$784 & 12.24 & 63.65 \\
		4 & [10\%, 15\%)    &  87$|$784 & 11.10 & 89.29 & 4 & [10\%, 15\%)        & 134$|$784 & 17.09 & 80.74 \\
		5 & [15\%, 20\%)    &  43$|$784 &  5.48 & 94.77 & 5 & [15\%, 20\%)        &  66$|$784 &  8.42 & 89.16 \\
		6 & [20\%, 35\%)    &  31$|$784 &  3.95 & 98.72 & 6 & [20\%, 35\%)        &  39$|$784 &  4.97 & 94.13 \\
		7 & [35\%, 50\%)    &  10$|$784 &  1.28 & 100.00 & 7 & [35\%, 50\%)        &  24$|$784 &  3.06 & 97.19 \\
		8 & [50\%, $+\infty$) &  0$|$784 &  0.00 & 100.00 & 8 & [50\%, $+\infty$)   &  22$|$784 &  2.81 & 100.00 \\
		\bottomrule
	\end{tabular}
	\caption{Ratio, percentage (PCT\%) and cumulative percentage (PCT.sum\%) of each specified relative difference range between the average AMR-based and MMR-based within standard deviation (denote as $\frac{\sum_{i=2}^{10} \sigma_{w.Ai} }{9}$ and $\frac{\sum_{i=2}^{10} \sigma_{w.Mi} }{9}$, for moving ranges sizes of 2 to 10) and the overall standard deviation ($\sigma_{overall}$), across 784 dimensions, showing the extent of deviation as a proportion of $\sigma_{overall}$.}
	\label{tab:ppk_cpk_ratio}
\end{table*}

The table shows that 388 out of 784 dimensions (49.49\%) have an absolute relative error between the average AMR-based standard deviation and the overall standard deviation is within 5\%. This increases to 78.19\% of dimensions within a 10\% error, and 94.77\% within a 20\% error. Such discrepancies can significantly impact PCIs, potentially leading to incorrect decisions if an inappropriate method is used to calculate standard deviation for a given scenario. Similarly, the right half of the T
able \ref{tab:ppk_cpk_ratio} presents the results for the MMR-based method.

\section{Simplified PCIs Workflow}
\begin{figure}[htbp]
	\centerline{\includegraphics[width=1.00\linewidth]{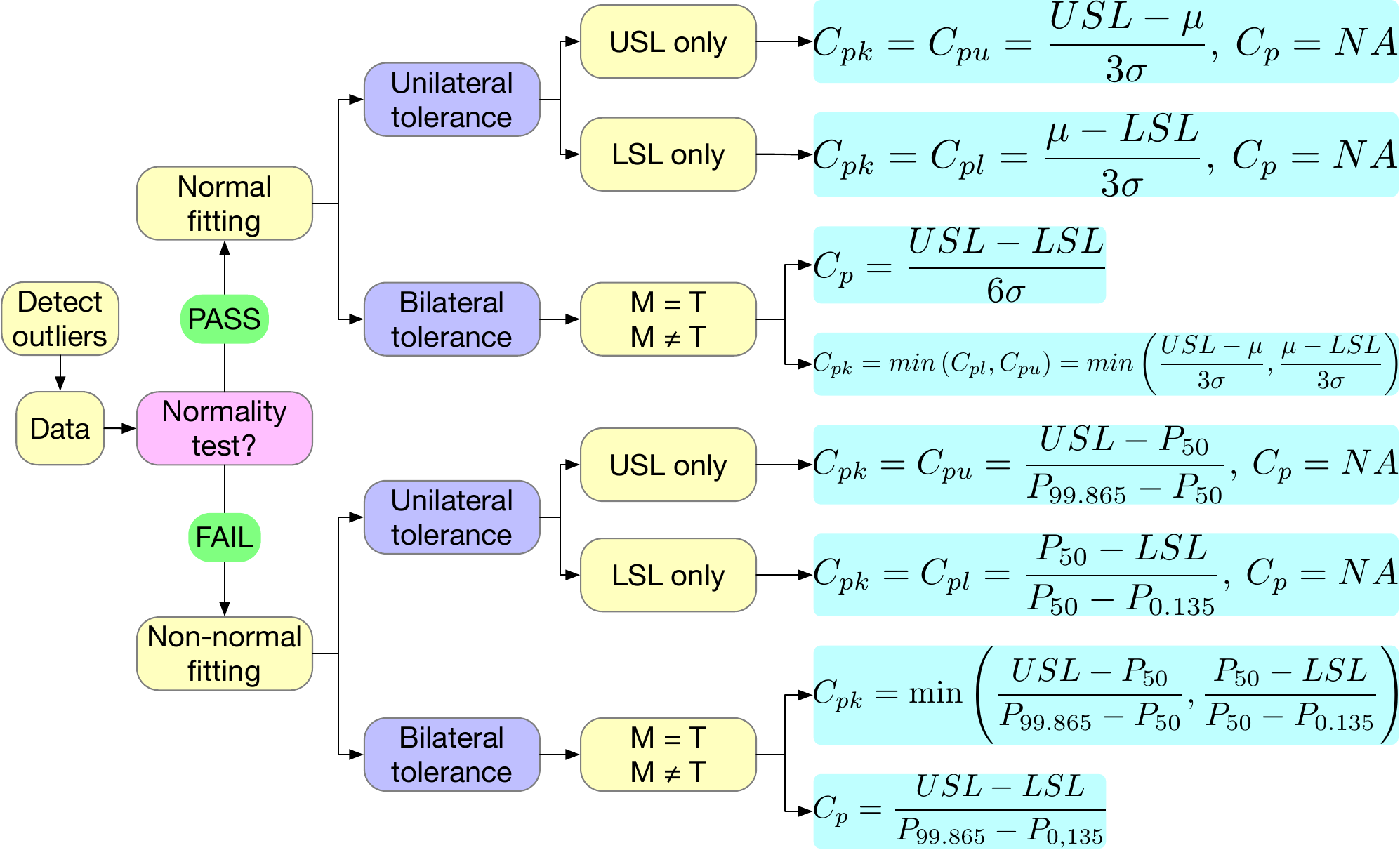}}
	\caption{PCIs workflow} 
	\label{pcis}
\end{figure}

Although the PCI workflow presented in Figure \ref{cpkwf} addresses most common scenarios, certain indices, such as $C_{pm}$ and $C_{pmk}$ are often perceived as overly complex or are inconsistently implemented in practice. As a result, the workflow is not universally adopted by researchers or engineers. To improve practicality and usability, a simplified version is shown in Figure \ref{pcis}, which excludes $C_{pm}$, $C_{pmk}$, and bilateral asymmetric tolerances. This simplification aligns with common manufacturing practices, where processes are typically centered within the specification limits unless otherwise specified by design requirements.

This simplified workflow highlights two key enhancements: outlier detection and non-normal distribution fitting. The outlier detection step identifies data points that may distort the statistical analysis or indicate root-cause dimensions. If the data fails a normality test, the non-normal fitting step selects the most appropriate distribution. This fitted distribution is then used to calculate process capability indices (PCIs) using non-normal methods. In addition, complementary visualization and statistical techniques will be employed to enhance data validation and analysis. These include the use of tolerance intervals, prediction intervals, and failure rate assessments expressed in parts per million ($PPM$).

\section{CONCLUSIONS}
This case study highlights the considerable variation in Process Capability Indices (PCIs) that arises solely from using different methods to calculate standard deviation. Notably, this variation occurs even before accounting for other critical factors, such as the type of tolerance (unilateral or bilateral, with or without a defined target), the nature of the data distribution (normal or non-normal), and the selected distribution fitting method. These findings underscore the sensitivity of PCI outcomes to the statistical assumptions and computational techniques employed. The use of inappropriate or inconsistent methods can lead to inaccurate assessments of process capability, potentially undermining quality control efforts. Therefore, the selection of a statistically robust and context-appropriate approach to PCI calculation is not merely a technical detail, but a crucial factor in ensuring reliable evaluations, sustaining process stability, and upholding effective quality management systems.

\section*{DECLARATION}
\begin{description}
	\item[Funding:] Not applicable. This research did not receive any specific grant from funding agencies in the public, commercial, or not-for-profit sectors.
	\item[Conflicts of interest / Competing interests:] The authors declare that they have no conflicts of interest or competing interests.
	\item[Availability of data and material:] Not applicable.
	\item[Code availability:] Not applicable.
	\item[Ethics approval:] Not applicable.
	\item[Consent to participate:] Not applicable.
	\item[Consent for publication:] All authors consent to the publication of this work and approve the final version of the manuscript.
\end{description}

\bibliographystyle{IEEEtran}
\bibliography{PCIs_bib}

\end{document}